\documentclass[twocolumn]{aastex62}
\usepackage{graphics,epsf}
\usepackage{amsmath}                % American Mathematical Society package
\usepackage{amsfonts}               % American Mathematical Society fonts
\usepackage{amssymb}                % American Mathematical Society symbol
\usepackage{epsfig}                 % EPS figures
\usepackage{graphicx}
\usepackage{float}
\usepackage{color}
\usepackage[para,online,flushleft]{threeparttable}

\newcommand{\kms}{{~\rm km\; s^{-1}}}

\newcommand{\cm}{{~\rm cm}}
\newcommand{\km}{{~\rm km}}
\newcommand{\s}{{~\rm s}}

\newcommand{\g}{{~\rm g}}

\newcommand{\erg}{{~\rm erg}}

\newcommand{\ms}{{~\rm ms}}

\begin{document}

\title{Low energy core collapse supernovae in the frame of the jittering jets explosion mechanism}

\author{Roni Anna Gofman}
\affiliation{Department of Physics, Technion, Haifa, 3200003, Israel; rongof@campus.technion.ac.il; soker@physics.technion.ac.il}

\author[0000-0003-0375-8987]{Noam Soker}
\affiliation{Department of Physics, Technion, Haifa, 3200003, Israel; rongof@campus.technion.ac.il; soker@physics.technion.ac.il}
\affiliation{Guangdong Technion Israel Institute of Technology, Shantou 515069, Guangdong Province, China}

\begin{abstract}
We relate the pre-explosion binding energy of the ejecta of core-collapse supernovae (CCSNe) of stars with masses in the lower range of CCSNe and the location of the convection zones in the pre-collapse core of these stars, to explosion properties in the frame of the jittering jets explosion mechanism. Our main conclusion is that in the frame of the jittering jets explosion mechanism the remnant of a pulsar in these low energy CCSNe has some significance, in that the launching of jets by the newly born neutron star (NS) spins-up the NS and create a pulsar. 
We crudely estimated the period of the pulsars to be tens of milliseconds in these cases.
The convective zones seed perturbations that lead to accretion of stochastic angular momentum that in turn is assumed to launch jittering jets in this explosion mechanism. We calculate the binding energy and the location of the convective zones with the stellar evolution code \textsc{mesa}. For the lowest stellar masses, we study, $M_{\rm ZAMS} \simeq 8.5-11 M_\odot$, the binding energy above the convective zones is low, and so is the expected explosion energy in the jittering jets explosion mechanism that works in a negative feedback cycle. The expected mass of the NS remnant is $M_{\rm NS} \approx 1.25M_\odot-1.6M_\odot$, even for these low energy CCSNe.
\end{abstract}


% ==========================================================
\section{INTRODUCTION}
\label{sec:intro}
% ==========================================================

In core-collapse supernovae (CCSNe) part of the core, or all of it and even part of the envelope, collapses to form a neutron star (NS), or a black hole, respectively. This process releases a huge amount of gravitational energy, most of which is carried by neutrinos, while a small fraction of the energy ejects the rest of the star. 
 
There are two basic processes that, in principle, might deliver part of the gravitational energy of the collapsing core to the exploding gas, the ejecta. One process is heating of the in-flowing gas by neutrinos, where the most commonly studied is the delayed neutrino mechanism \citep{BetheWilson1985}.
In the other process, jets that the newly born NS (or black hole) launches, even when the net angular momentum is very low, deliver the energy to the ejecta, this is the jittering jets explosion mechanism (e.g., \citealt{Soker2010, GilkisSoker2014, Quataertetal2019}). The jets operate in a negative feedback mechanism \citep{Soker2016Rev}. Namely, when the jets manage to eject the outer parts of the core (or of the envelope in the case of black hole formation) accretions stops, and so are the jets.

Studies in recent years found that each of these two mechanisms requires one or more additional ingredients for a successful explosion. The problems of the delayed neutrino mechanism (e.g., \citealt{Papishetal2015, Kushnir2015b}) brought people who conduct CCSN simulations to introduce convection in the pre-collapse core (e.g., \citealt{CouchOtt2013, Mulleretal2019Jittering}). These convective flow fluctuations lead to relatively large stochastic angular momentum variations of the flow onto the newly born NS. Some simulations show that these stochastic variations result in the formation of jittering jets, namely, the axis of a bipolar outflow changes its direction \citep{Soker2019JitSim}. The claim is that the extra ingredient that the delayed neutrino mechanism requires might be jittering jets (e.g., \citealt{Soker2019JitSim}). 

The recent study by \cite{SawadaMaeda2019} seems to support this claim. In most of the simulations that do reach an explosion, the process is slow, reaching the explosion energy in a time of $t_{\rm exp}> 1 \s$.  \cite{SawadaMaeda2019} argue that nucleosynthesis yields require explosion on a time scale of $t_{\rm exp} \la 0.25 \s$, this brings them to ``... suggest that there must be a key ingredient still missing in the ab-initio simulations, which should lead to the rapid explosion.''
 We take the view that this ingredient is the process of jittering jets. 

Convective flow fluctuations in the pre-collapse core (in the case of NS formation) or envelope (in the case of black hole formation) are the base of the jittering jets explosion mechanism \citep{GilkisSoker2014, GilkisSoker2015, Quataertetal2019}. Another key process is the amplification of the initial fluctuations by instabilities behind the shock of the inflowing gas (the stalled shock) at $r \simeq 100 \km$ from the centre, e.g., the spiral standing accretion shock instability (SASI; e.g., \citealt{BlondinMezzacappa2007, Iwakamietal2014, Kurodaetal2014, Fernandez2015, Kazeronietal2017}, for studies of the SASI). The jittering jets explosion mechanism assumes that the angular momentum fluctuations bring the newly born NS (or black hole) to launch jittering jets.
However, numerical simulations (yet) find no stochastic accretion disks around the newly born NS. This might suggest that the jittering jets explosion mechanism also require an extra ingredient, e.g., heating by neutrinos \citep{Soker2018KeyRoleB, Soker2019SASI, Soker2019JitSim}.

All these studies lead to a picture where both jittering jets and neutrinos heating play some roles. The question is which of the two dominate. 
 
There are several cases where the delayed neutrino mechanism and the jittering jets explosion mechanism predict different outcomes. For example, according to the delayed neutrino mechanism, most stars with a zero-age main sequence (ZAMS) mass of $M_{\rm ZAMS} \ga 18 M_\odot$ do not explode, but rather lead to a failed SN and the formation of a black hole (e.g., \citealt{Fryer1999, Ertletal2020}). 
The jittering jets explosion mechanism predicts that all stars explode, even when they form a black hole, as even the black hole launches jittering jets that lead to an explosion  (e.g., \citealt{GilkisSoker2014, GilkisSoker2015, Quataertetal2019}). In the jittering jets explosion mechanism, there are no failed SNe, but rather the formation of a black hole might lead to energetic explosions with energies up to $E_{\rm exp} > 10^{52} \erg$ \citep{Gilkisetal2016Super}. In a previous paper \citep{Gofmanetal2019}, we study these differences.

In the present study, we examine the differences between the two explosion mechanisms in the case of the lower mass range of CCSNe, $8.5 \la M_{\rm ZAMS} \la 11.5 M_\odot$, concerning low-energy CCSNe.
We review some properties of low-energy CCSNe with Pulsars in section 
\ref{sec:properties}. When then describe the numerical method in section \ref{sec:Numerical}, the relevant stellar properties in section \ref{sec:StellarProperties}, and the implicaitons in section \ref{sec:Implicaitons}. We summarise our main results in section \ref{sec:summary}.

% ==========================================================
\section{The relevant properties of supernova remnants with pulsars}
\label{sec:properties}
% ==========================================================
There are several CCSNe with very low explosion energy of $E_{\exp} \approx 10^{49} - 10^{50} \erg$. These SN remnants (SNRs) contain a pulsar with a pulsar wind nebula (PWN; see, e.g., \cite{Martin2014}). A famous case is the Crab Nebula with an explosion energy of  $E_{\rm exp} \approx 10^{50} \erg$ (e.g., \citealt{YangChevalier2015}).
\cite{ReynoldsBorkowski2019} estimate the explosion energy of the G310.6−1.6 SNR that has a PWN to be very low, $E_{\rm exp} \approx 3 \times 10^{47} \erg$, and the ejected mass to be $M_{\rm eject} \approx 0.02 M_\odot$. This very low energy CCSN is a puzzle. 
\cite{Guestetal2019} estimate the kinetic energy, which is about the explosion energy, of the SNR~G21.5−0.9 to be $E_{\rm exp} \simeq 3 \times 10^{49} \erg$.
\cite{Temimetal2019} study the SNR/PWN Kes~75 and conclude that the progenitor had a mass of $M_{\rm ZAMS} = 8 - 12 M_\odot$. They model this SNR with an explosion energy of $E_{\rm exp} \simeq 6 \times 10^{50} \erg$. 
We note that some low energy CCSNe might be electron capture CCSNe (.e.g, \citealt{Nomotoetal2014}). For that, we follow the oxygen core in the lower mass range of CCSNe.   

The difference between the two explosion mechanisms concerning these CCSNe is that in the delayed neutrino explosion mechanism the presence of a pulsar, as opposed to a non-magnetic non-rotating NS, is of no significance, while in the jittering jets explosion mechanism the rotation and magnetic fields are the results of the launching of jittering jets by the newly born NS and the region around it.  

The jittering jets explosion mechanism can account for the low explosion energy and relatively slow pulsar rotation at birth, i.e., much below break-velocity  (we discuss this point in Sec. \ref{subsec:NSspin}).
\cite{vanderSwaluwWu2001} find the typical spin period at birth to be $\tau_s \approx 40 \ms$ with a large scatter.  
The low explosion energy comes from the low binding energy of the ejecta. The accretion of gas with small amounts of angular momentum accounts for the slow rotation, while the large angular momentum fluctuations allow the launching of a jittering bipolar outflow (jittering jets; \citealt{Soker2019JitSim} and references therein). 

\cite{OzelFreire2016} review the masses of NS in binary systems, and find the mass distribution to be wide, $1 M_\odot \la M_{\rm NS} \la 2 M_\odot$. \cite{Tangetal2019} use the NS equation of state to infer the masses of three isolated NSs. This method has much larger uncertainties than the methods for binary systems, and \cite{Tangetal2019} deduce a wide mass distribution for each one of the three. Crudely, the three NSs mass distribution is in the range of $M_{\rm NS} \approx 0.9 - 1.5 M_\odot$. The jittering jets explosion mechanism is sensitive to four parameters of the pre-collapse star: The angular momentum of the core (rotation velocity), the fluctuating convective flow in the core, the magnetic field in the core, and the binding energy of the ejected mass \citep{Soker2018KeyRoleB, Soker2019SASI, Soker2019JitSim}. These sensitivities explain the wide NS mass distribution. 
We do note that since the jittering jets explosion mechanism is sensitive to the angular momentum of the collapsing core, the outcome of an exploding star in a close binary system might be different, in energy and explosion morphology, than that of a single star.

In what follows, we examine the binding energy and the location of the convective zones of the pre-collapse cores. We do not consider angular momentum and magnetic fields, and therefore our study is limited in its implications. Nonetheless, we can still shed light on the expected outcome of the jittering jets exploding mechanism of low mass stars.  

% ==========================================================
\section{Numerical Set Up}
\label{sec:Numerical}
% ========================================================== 

We evolve single stellar models with ZAMS mass in the range of $8.5M_\odot-15M_\odot$  and  ZAMS metallicity of $Z=0.02$ using the Modules of Experiments in Stellar Astrophysics code (\textsc{mesa}, version 10398 \citealt{Paxton2011,Paxton2013,Paxton2015,Paxton2018}).  We evolve the stellar models from pre-main sequence up to one of three evolutionary stages as follows: (1) $\rm{Si_{nuc}}$ - oxygen is burning to create silicon, namely the total power from all nuclear reactions is high than $10^{10} L_\odot$; (2) $\rm {Si_{core}}$ - a silicon core has formed and has mass larger than $1.5M_\odot$; or (3) $\rm {Fe_{core}}$ - an iron core has formed and has began to collapse, namely the infall velocity is high than $1000 \km \s^{-1} $.

We use the \textsc{mesa} "Dutch" scheme for massive stars wind mass-loss. This scheme combines results from several paper and is based on \cite{Glebbeek2009} and is as follows. For stars with effective temperature, $T_{\rm eff} > 10^4 ~{\rm K}$ and surface hydrogen abundance, $X_{\rm s}$, above $0.4$ we use \cite{Vink2001}, and for such hot stars with $X_{\rm s}<0.4$ we use \cite{WindWR}. In cases where $T_{\rm eff} < 10^4 ~{\rm K}$, we apply mass loss according to \cite{deJager1988}.  We set the mass-loss scaling factor to be $0.8$ since we assume that the stars do not rotate  \citep{Maeder2001}. 

We employ mixing in convective regions defined by the Ledoux criterion according to a mixing-length theory \citep{Henyey1965} with $\alpha_{\rm MLT}=1.5$ and $\alpha_{\rm sc}=1.0$ for semiconvection \citep{Langer1983}. We apply convective overshooting using a step function with an overshooting parameter of $0.335$ \citep{Brott2011}.

% ==================================
\section{ Stellar properties} 
\label{sec:StellarProperties}
% ==================================
 When presenting our results, we concentrate on properties that are relevant to the jittering jets explosion mechanism (Sec. \ref{sec:intro}) concerning low-energy CCSNe with pulsar remnants (Sec. \ref{sec:properties}). The two main properties are the pre-collapse binding energy of the mass ejected in the explosion and the perturbation due to convection in the pre-collapse core.  Instabilities in the post-shock region above the NS amplify these perturbations (e.g., \citealt{KazeroniAbdikamalov2020} for a recent paper), including large-amplitude variations in angular momentum of the gas accreted onto the NS (Sec.  \ref{sec:intro}). The binding energy is relevant to the explosion energy because the jittering jets explosion mechanism operates in negative feedback, this implies that the explosion energy is about the binding energy for a high-efficiency explosion, and several times the binding energy for low efficiency (section \ref{sec:intro}).

We calculate the binding energy of the outer part of the core and the entire envelope, from a mass coordinate $M_{\rm in}$ in the core to the stellar surface.  The binding energy $E_{\rm bind}$ includes the internal energy of the gas and the gravitational energy, and \textsc{mesa} supplied it.

\begin{figure}
\centering
\includegraphics[trim= 0cm 0cm 0cm 0cm,clip=true,width=0.47\textwidth] {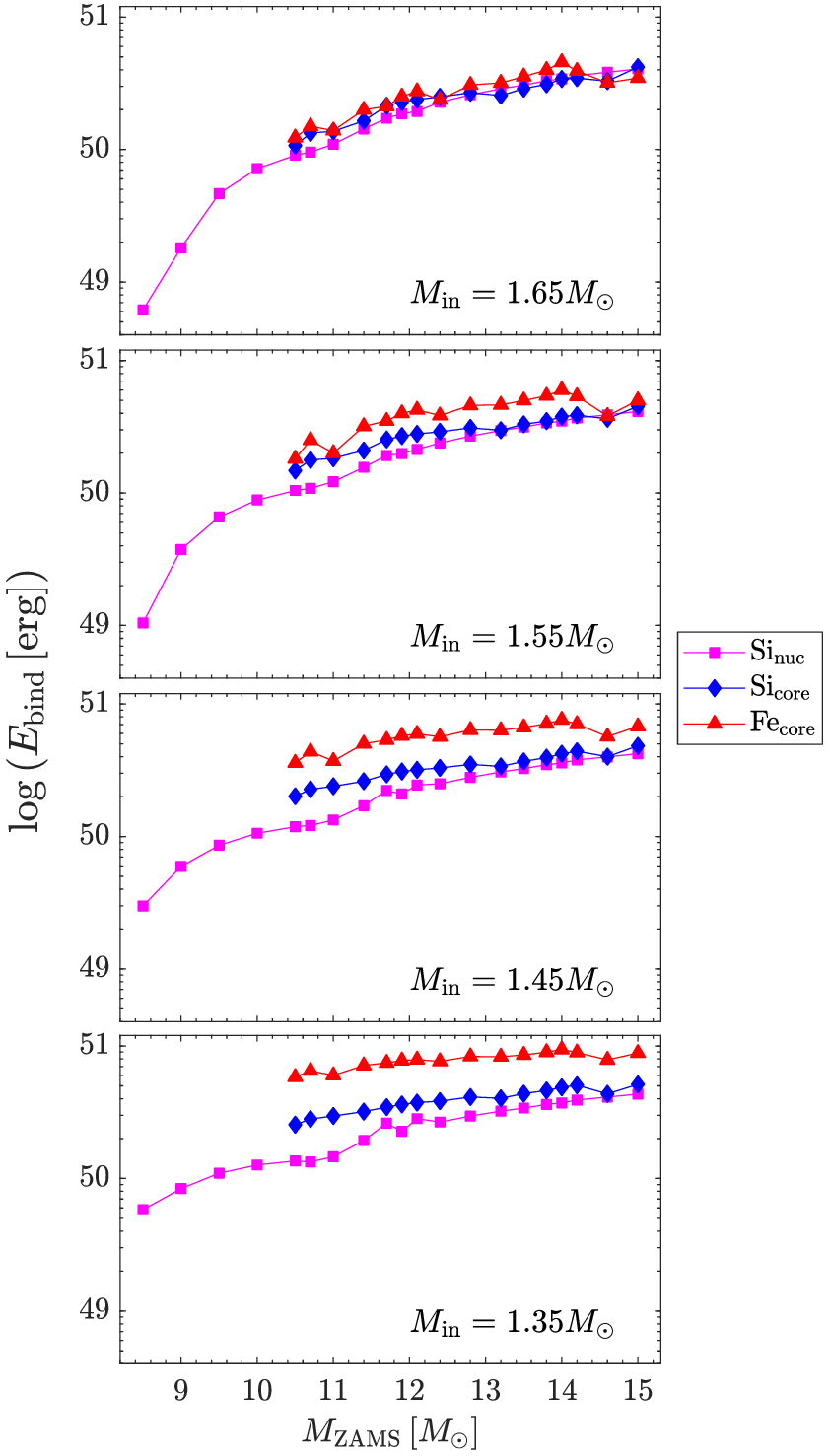} 
\caption{The binding energy integrated from 4 different mass coordinate $M_{\rm in}$ to $M_*$ as a function of the ZAMS mass. Models marked with magenta squares evolved up to oxygen burning ($\rm{Si_{nuc}}$), models marked with blue diamond evolved up to silicon core formation ($\rm{Si_{core}}$), and models marked with red triangles evolved up to iron core formation ($\rm {Fe_{core}}$).   }
\label{fig:Ebind}
\end{figure}

In Fig. \ref{fig:Ebind} we present binding energies for different initial masses and for models at the three evolutionary stages as follows (see Sec. \ref{sec:Numerical}). (1) $\rm{Si_{nuc}}$: Oxygen burning after production of some silicon. (2) $\rm{Si_{core}}$: Just before the silicon starts to burn and there is a massive silicon core. (3) $\rm{Fe_{core}}$: Just before collapse when there is a large iron core. Due to numerical problems, we could not run some low mass models to the phase of silicon burning.

% FFFFFFFFFFFFFFFFFFFFFFFFFFFFFFFFFFFFFFFFFFFFFFFFFFFFFFFFFFFFFFFFFFFF
% FFFFFFFFFFFFFFFFFFFFFFFFFFFFFFFFFFFFFFFFFFFFFFFFFFFFFFFFFFFFFFFFFFFF

We learn from the upper panel of Fig. \ref{fig:Ebind} that for the mass coordinate $M_{\rm in} =1.65M_\odot$ the binding energy for all three stages is approximately similar for stars with $M_{\rm ZAMS}>10.5M_\odot$, namely, the binding energy above this mass coordinate does not change much during the formation of
a silicon core and then an iron core.
We extended this conclusion to include stars with $M_{\rm ZAMS} <10.5M_\odot$, and assume that their binding energy above $M_{\rm in} =1.65M_\odot$ at core-collapse is similar to that at $\rm{Si_{nuc}}$ stage.
We can assume the same, but with more significant differences between binding energies at the three different stages, for the mass coordinate $M_{\rm in}=1.55M_\odot$. We comment below on those that will explode as electron capture supernovae and will not reach a silicon core. 

We note that for stars with $M_{\rm ZAMS} \la 10 M_\odot$ electron capture before full oxygen burning might trigger core-collapse (e.g., \citealt{LeungNomoto2019}). In these cases, the binding energy at the stage ${\rm Si_{nuc}}$ is a good approximation to the relevant binding energy at the explosion.

Taking these assumptions into account, stars with $M_{\rm ZAMS} \lesssim 11 M_\odot$ have rather low binding energies, $6 \times 10^{48} \erg \lesssim E_{\rm bind} \lesssim 10^{50} \erg$ above the mass coordinates $M_{\rm in}=1.65M_\odot$, and $10^{49} \erg \lesssim E_{\rm bind} \lesssim 2 \times 10^{50} \erg$ above the mass coordinates $M_{\rm in}=1.55M_\odot$. This is in line with earlier results, (e.g., \citealt{Sukhboldetal2016}) that for $M_{\rm ZAMS} \lesssim 10 M_\odot$ the core is less dense, and hence binding energy decreases with decreasing $M_{\rm ZAMS}$.

\begin{figure*}
\centering
\includegraphics[trim= 0cm 0cm 0cm 0cm,clip=true,width=0.97\textwidth] {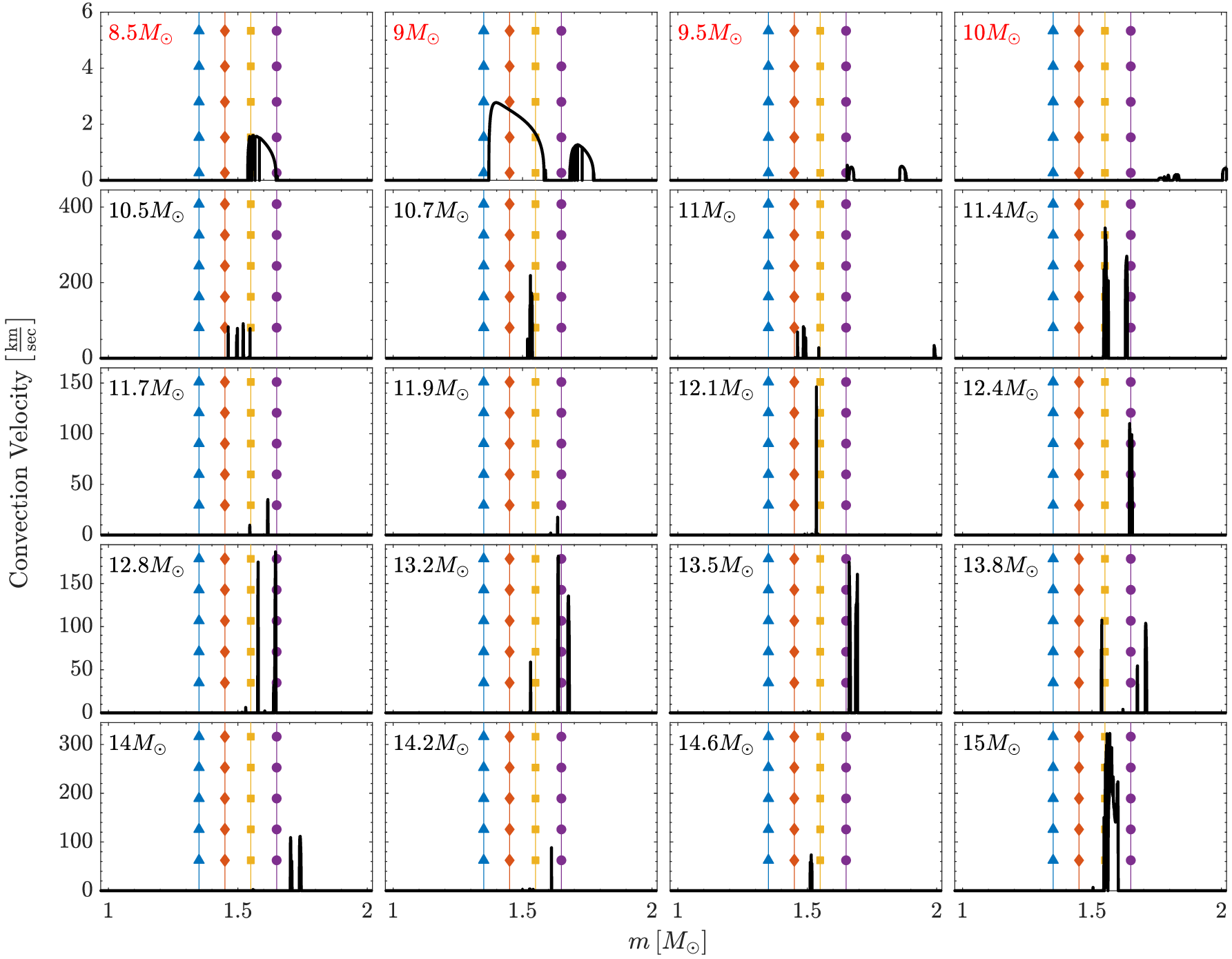} 
\caption{The convection velocity in $\kms$ as a function of the mass coordinate at the sage of $\rm {Si_{nuc}}$ for models with $M_{\rm ZAMS} < 10.5 M_\odot$ (red mass label) and at the $\rm {Fe_{core}}$ stage for models with $M_{\rm ZAMS} \ge 10.5 M_\odot$ (black mass label). The 4 vertical lines are the 4 mass coordinates, $M_{\rm in}$, for which we calculated envelope binding energy in Fig. \ref{fig:Ebind}: $1.35M_\odot$ in blue triangles, $1.45M_\odot$ in orange diamonds, $1.55M_\odot$ in yellow squares and $1.65M_\odot$ in purple circles.}
\label{fig:ConvVeM}
\end{figure*}

Fig. \ref{fig:ConvVeM} shows the convection velocity for the models we presented in Fig. \ref{fig:Ebind} at the sage of $\rm {Si_{nuc}}$ for models with $M_{\rm ZAMS} < 10.5 M_\odot$ and at the $\rm {Fe_{core}}$ stage for models with $M_{\rm ZAMS} \ge 10.5 M_\odot$.
% FFFFFFFFFFFFFFFFFFFFFFFFFFFFFFFFFFFFFFFFFFFFFFFFFFFFFFFFFFFFFFFF
% FFFFFFFFFFFFFFFFFFFFFFFFFFFFFFFFFFFFFFFFFFFFFFFFFFFFFFFFFFFFFFFF

\begin{figure}
\centering
\includegraphics[trim= 0cm 0cm 0cm 0cm,clip=true,width=0.4\textwidth] {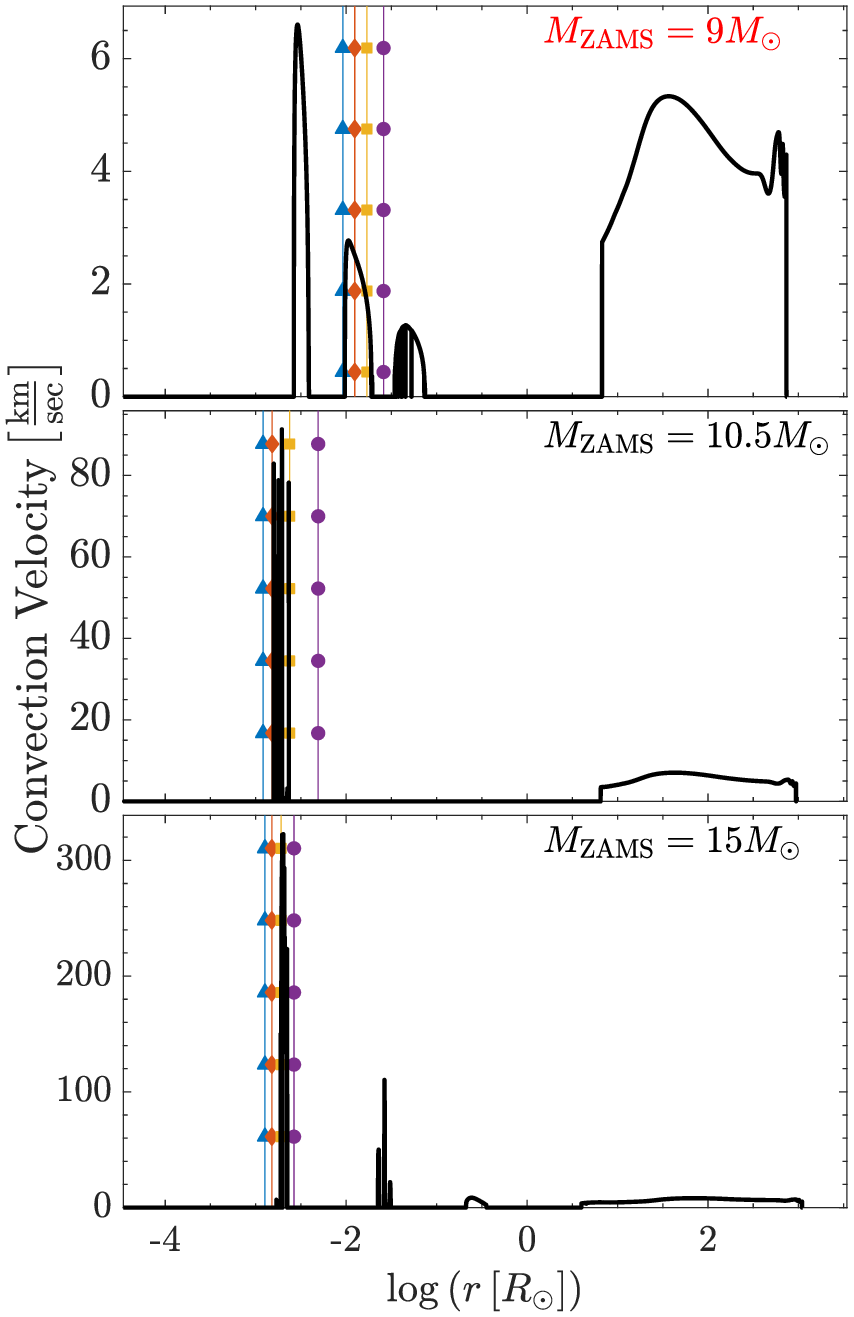} 
\caption{The convective velocity in $\kms$ as a function of the radius in a logarithmic scale at the $\rm{Si_{nuc}}$ stage for the $M_{\rm ZAMS} < 9.5 M_\odot$ model, and at the $\rm{Fe_{core}}$ stage for the two lower models.  The 4 vertical lines have the same meaning as in Fig. \ref{fig:ConvVeM}.  }
\label{fig:ConvVeR}
\end{figure}

In Fig. \ref{fig:ConvVeR} we show the convective velocity as function of radius for three models at the end of the simulation. From this figure we learn that in the low mass models that are likely to end as electron capture CCSNe, the relevant mass coordinate (marked with vertical lines) occur at larger radii. Therefore, although the convective velocity is low, the stochastically fluctuating angular momentum is non-negligible. 
% FFFFFFFFFFFFFFFFFFFFFFFFFFFFFFFFFFFFFFFFFFFFFFFFFFFFFFFFFFFFFFFF
% FFFFFFFFFFFFFFFFFFFFFFFFFFFFFFFFFFFFFFFFFFFFFFFFFFFFFFFFFFFFFFFF

% ===================================================
\section{Implications}
\label{sec:Implicaitons}
% ===================================================

Although most of the findings we present in section \ref{sec:StellarProperties} are not new by themselves, the way we present them, namely, binding energy and convective velocities, allows us to relate these properties to the formation of low energy CCSNe with pulsars to the jittering jets explosion mechanism. Namely, to relate the binding energy and location of convective velocities to the properties of the explosion and of the NS remnant. These properties are the explosion energy, NS mass, and  
NS spin period. 
 
% =======================
\subsection{Explosion Energy and NS Mass}
\label{subsec:Energy}
% =======================

In the jittering jets explosion mechanism the accretion of gas with stochastic angular momentum onto the newly born NS is crucial. The stochastic angular momentum starts with convective fluctuations in the core (Sec. \ref{sec:intro}), namely the convective velocities as we present in Fig. \ref{fig:ConvVeM}. For low mass stars that explode by electron capture the convective velocity is lower, but occurs at larger distances from the center (Fig. \ref{fig:ConvVeR}). The assumption is that the stochastic accretion of angular momentum lead to the launching of jets with the aid of neutrino heating (Sec. \ref{sec:intro}). When the jets manage to eject the core, accretion stops (a negative feedback). As more mass is accreted, the energy of the jets increases, and the binding energy of the remaining mass decreases, as we see in Fig. \ref{fig:Ebind}.  

From Fig \ref{fig:ConvVeM} we learn that the convective regions occur at mass coordinate of $m \simeq 1.35-1.7 M_\odot$. These will seed fluctuations above the newly born NS that will lead to jittering jets. Such a baryonic mass forms a NS of mass $M_{\rm NS} \simeq m-0.1 M_\odot \simeq 1.25-1.6 M_\odot$ \citep{Sukhboldetal2016}. The explosion energy is the energy that the jets carry minus the binding energy. If we assume no fine tuning, we expect the explosion energy to be $E_{\exp} \approx {\rm few} \times 0.1 E_{\rm bind} - {\rm few} \times E_{\rm bind}$.  
   
From Figs. \ref{fig:Ebind} and \ref{fig:ConvVeM} the explosion energy of most stars with mass in the range of $8 \la M_{\rm ZAMS} \la 10 M_\odot$, in the jittering jets explosion mechanism to be in the range of $E_{\rm exp} \approx {\rm several} \times 10^{48} \erg - 10^{50} \erg$. For most stars in the range $10 \la M_{\rm ZAMS} \la 12 M_\odot$ we expect the jittering jets explosion energy to be 
$E_{\rm exp} \approx 10^{50} - 10^{51} \erg$.
In some cases the efficiency might be low, and explosion energy large, even much larger than the binding energy. This might be the case when the pre-collapse core is rapidly rotating and the jets are well collimated and break out along the fixed polar directions \citep{Gilkisetal2016Super}.  
 
% =======================
\subsection{NS Spin Period}
\label{subsec:NSspin}
% =======================

We consider here a slowly rotating pre-collapse core (practically zero rotation velocity). There are local angular momentum fluctuations in the convective zones of the pre-collapse core (Fig. \ref{fig:ConvVeM}). The key process in launching jittering jets is that instabilities in the post-shock region ($r \la 100 \km$), mainly the spiral-SASI (e.g., \citealt{BlondinMezzacappa2007, Kazeronietal2017}; see Sec. \ref{sec:intro}), amplify these perturbations  by a large factor. It is impossible to conduct 3D hydrodynamical simulations to follow the entire process of core collapse, mass accretion onto the newly born NS and, most important, the launching of jets. We bypass these uncertainties by directly estimating the angular momentum that the jets are likely to carry away. The NS will have at the end of the process an opposite angular momentum. 
 
We consider some jet's properties that might be common to CCSNe and other systems that launch jets, i.e., terminal velocities about equal to the escape velocity, and specific angular momentum that is about few times the Keplerian one at the launching radius (see review by \citealt{Soker2016Rev}). Using these and more assumptions as we discuss next,
we can crudely estimate the angular momentum that the jets carry. As said, because we assume that the initial core angular momentum is very small, an opposite angular momentum is `left' in the newly born NS (pulsar). 

We make the following assumptions and definitions. 
(1) We assume that the jets remove the ejecta with an efficiency $\eta_{\rm eff}\equiv E_{\rm bind}/ E_{\rm jets} <1$, so that the explosion energy is 
\begin{equation}
E_{\rm exp} = E_{\rm jets}- E_{\rm bind} =  E_{\rm jets} \left(1-\eta_{\rm eff} \right),  
\label{eq:Eexp}
\end{equation}
where $E_{\rm jets}$ the jets energy, and $\eta_{\rm eff} \approx 0.5$, i.e.,  $E_{\rm jets} \approx 2 E_{\rm bind}$.

We base the scaling of $\eta_{\rm eff} \approx 0.5$ on the results of \cite{Gilkisetal2016Super}, who concluded that for regular CCSNe, namely those that are not super-energetic, the efficiency of mass removal by jets should be larger than about 0.4 (for their $12 M_\odot$ stellar model; their figure 4). Although they use mass removal efficiency and we use energy efficiency, we take the energy efficiency to be large as well. Namely, a jets' energy only twice the binding energy is sufficient to explode the star.  
This is reasonable for these low energy CCSNe where according to the jittering jets explosion mechanism the feedback cycle is more efficient \citep{Gilkisetal2016Super}. 
Another reason for this approximate scaling is the following. For typical explosion energies of $\approx 10^{51} \erg$, \cite{Gilkisetal2016Super} estimate (based on earlier hydrodynamical simulations by, e.g, \citealt{PapishSoker2014}) that the jets energy is $E_{\rm jets}({\rm typical ~CCSNe}) \approx (3-5) \times E_{\rm bind} \approx 10^{51} \erg$. Therefore, for a more efficient jet feedback cycle we take $E_{\rm jets} \approx 2 E_{\rm bind}$.
We bring another derivation of the relation between the explosion energy and the binding energy in Appendix \ref{AppendixA}.  
  
(2) We assume that the jets have a terminal specific energy, mostly kinetic energy, about equal to the specific escape energy from from their average launching radius $r_{\rm jets}$. This gives a terminal velocity of about the escape speed, $v_{\rm jets} \simeq v_{\rm esc}=\sqrt{2 G M_{\rm NS}/ r_{\rm jets}}$, such that the mass carried by the jets is
\begin{equation}
M_{\rm jets} = \frac{2E_{\rm jets}}{v_{\rm jets}^2} 
     = \frac{ E_{\rm exp}  r_{\rm jets} }{G M_{\rm NS}} \left(1-\eta_{\rm eff}\right)^{-1}.
\label{eq:Mjets}
\end{equation}

(3) We assume that the specific angular momentum of the jet is about few times that at its average launching radius $j_{\rm jets} = \beta_{j} (G M_{\rm NS} r_{\rm jets})^{1/2}$.
In many jet-launching models the specific angular momentum of the matter in the jet is few times that at the launching radius, i.e., $\beta_{j} \simeq 3$ (e.g., \citealt{Andersonetal2003}).
  
(4) We take $N$ stochastic jet-launching episodes. 
   
(5)  We assume that the jets are launched from the zone where the SASI occurs, as in one of the jet-launching mechanism for the jittering jets explosion mechanism when the angular momentum of the pre-collapse core is very low \citep{Soker2019SASI}. Namely, 
$r_{\rm jets} \simeq 50 \km$.

Substituting for the jet velocity we can write the total angular momentum that the jets carry as 
\begin{equation}
J_{\rm jets}= \frac{M_{\rm jets} j_{\rm jets}}{\sqrt{N}}  = \frac{E_{\rm exp} }{\sqrt{N}}
\frac{r^{3/2} _{\rm jets} \beta_{j}}{(G M_{\rm NS})^{1/2}} 
\left(1-\eta_{\rm eff} \right)^{-1}. 
\label{eq:Jtot1}
\end{equation}
If it is only due to this angular momentum that the NS spins, i.e., the pre-collapse core has zero angular momentum, then the spin period of the newly born NS is  
$\tau_s= 2 \pi I_{\rm NS}/J_{\rm jets}$. This reads 
\begin{equation}
\begin{aligned}
    \tau_s \simeq & \; 60  
    \left( \frac{\beta_{j}}{3} \right)^{-1}
    \left( \frac{1-\eta_{\rm eff}}{0.5} \right)
    \left( \frac{\sqrt{N}}{\sqrt{10}} \right) 
    \\  &  \times
    \left( \frac{E_{\rm exp}}{10^{50} \erg} \right)^{-1}
        \left( \frac {M_{\rm NS}}{1.4 M_\odot} \right)^{1/2} 
    \left( \frac{r_{\rm jets}}{50 \km} \right)^{-3/2}  
    \\  &  \times
    \left( \frac{I_{\rm NS}}{1.5 \times 10^{45} \g \cm^2} \right) \ms,  
\end{aligned}     
\label{eq:Jtot3}
\end{equation}
where we take the moment of inertia of the NS, $I_{\rm NS}$, from \cite{Worleyetal2008}. 
We note that the typical efficiency values that \cite{Gilkisetal2016Super} use for typical CCSNe (non super-energetic) correspond here to $\eta_{\rm eff} \approx 0.4-0.5$. Even if we consider the larger range of $\eta_{\rm eff} \approx 0.3-0.6$, the value of the spin period will not change much and be in the range of $\tau_s \approx 85 - 50 \ms$.

Equation (\ref{eq:Jtot3}) has several implications. We here mention two. The first implication is for the most energetic explosions, namely, CCSNe with explosion energies of $E_{\rm exp} \ga 10^{52} \erg$. In these cases, the spin period according to the jittering jets explosion mechanism reaches the shortest possible value of  $\tau_s \simeq 1 \ms$. If the NS magnetic field is strong enough, we have an energetic magnetar. Namely, energetic magnetars come along with energetic jets in the explosion \citep{Soker2016Magnetar, SokerGilkis2017Magnetar}.  The jets are likely to carry more energy than the magnetar. 

The other implication is relevant to our study of the least energetic CCSNe. For these CCSNe, the total angular momentum that the jets leave on the NS is very low, corresponding to $\tau_s \approx {\rm few} \times 10 - 100 \ms$. These values are compatible with the observed values of $\tau_s \approx 40 \ms$ \citep{vanderSwaluwWu2001}, and with new expectation of slow NS birth rotation \citep{Fulleretal2019}. 

% ==========================================================
\section{Summary}
\label{sec:summary}
% ==========================================================
 
We calculated the binding energy of the outer core and envelope, and located the convection regions in the core, as both properties are related to the jittering jets explosion mechanism. We related (Sec. \ref{sec:Implicaitons}) these to three observational properties of SNRs and their pulsar, i.e., the SN explosion energy, the pulsar mass, and the pulsar spin period.
Using the numerical stellar evolution code \textsc{mesa}, we evolved stellar models with a mass in the range of $8.5M_\odot \leq M_{\rm ZAMS} \leq 15M_\odot$, and we concentrated in the lower masses. We calculated the binding energy at three different evolution stages, during oxygen burning to silicon, after the formation of a massive silicon core, and just before core collapse when the inner part is iron. We presented the binding energies of mass residing above four mass coordinates in Fig. \ref{fig:Ebind}. 

We found that most models feature convective zones somewhere between at the mass coordinate $1.35M_\odot$ and $1.7M_\odot$ (Fig. \ref{fig:ConvVeM}). For low mass stars in our calculations where the convection velocity is slow, the convection region is at a larger radius  (Fig. \ref{fig:ConvVeR}), thus implying that the angular momentum fluctuations are of similar significance in all cases.
The convective regions seed the perturbations that lead to stochastic angular momentum accretion onto the newly born NS (or black hole), that are assumed to drive to the launching of jittering jets.
  
Though the calculations of binding energies and the locations of the convective zones are not new on their own, we have presented them together and associated them via the jittering jets explosion mechanism to the properties of the SNRs and their pulsar (Sec. \ref{sec:Implicaitons}). 
The jittering jets explosion mechanism works in a negative feedback cycle, such that when the jets manage to eject the core the accretion ceases, and so are the jets.  This implies that the explosion energy is of an order of magnitude of the binding energy of the pre-collapse outer core and envelope, namely the mass ejected at the explosion. 

For the lowest mass stars we studied the binding energy can be very low. From Fig. \ref{fig:Ebind} we conclude that the binding energy above the mass coordinates $M_{\rm in} = 1.65M_\odot$  and $M_{\rm in} =1.55M_\odot$ does not change much during the formation of a silicon core and later or an iron core. Stars with $M_{\rm ZAMS} \lesssim 11 M_\odot$ have rather low binding energies of $6 \times 10^{48} \erg \lesssim E_{\rm bind} \lesssim 10^{50} \erg$ above the mass coordinates $M_{\rm in}=1.65M_\odot$, and $10^{49} \erg \lesssim E_{\rm bind} \lesssim 2 \times 10^{50} \erg$ above the mass coordinates $M_{\rm in}=1.55M_\odot$.

The presence of convective zones in the mass coordinate range above which binding energy is low for the lowest mass stars we have studied implies for the jittering jets explosion mechanism that the mass of the newly born NS is $M_{\rm NS} \approx 1.25M_\odot-1.6M_\odot$, even for low energy CCSNe.

Our main conclusion is that in the frame of the jittering jets explosion mechanism, the remnant of a pulsar has some significance. The jets carry with them angular momentum, and therefore the opposite amount of angular momentum is left to spin-up the NS and create a pulsar. We crudely estimated the period of the pulsars (Sec. \ref{subsec:NSspin}) to be tens of milliseconds (Eq. \ref{eq:Jtot3}).

%\vspace{0.1cm}
% ==========================================================
\section*{Acknowledgements}
% ==========================================================
We thank an anonymous referee for helpful suggestions. This research was supported by a grant from the Israel Science Foundation.
%\vspace*{0.1cm}

%%%%%%%%%%%%%%%%%%%%%%%%%%%%%%%%%%%%%%%%%%%%%%%%
\appendix\section{Explosion energy and binding energy}
\label{AppendixA}
%%%%%%%%%%%%%%%%%%%%%%%%%%%%%%%%%%%%%%%%%%%%%%%%

We here consider the general nature of the feedback mechanism of the jittering jets explosion mechanism.

We follow \cite{PapishSoker2012} and present a simple spherically-symmetric expansion of a shock through the infalling material and the inner part of the core, up to $r \approx 10^4 \km$.  \cite{PapishSoker2012} approximate the hot bubble that the jittering jets form as a spherical hot bubble that pushes out from the core. 
We use the same density profile and the same equations as they used for a typical CCSN. This implies that we demonstrate the relation between explosion energy and binding energy for a typical CCSN, and not necessary for low energy ones. The scaling can be extended to lower or higher masses. 

We also follow \cite{PapishSoker2012} in their  assumptions and results of the jittering jets explosion mechanism (see their section 1). 
(1) The jets do not revive the stalled
shock, but rather penetrate through it.
(2) \cite{PapishSoker2011} showed that the jets penetrate the in-falling gas up to a distance of ${\rm few} \times 1000 \km$, that is, beyond the stalled shock at $\approx 100 \km$. However, beyond ${\rm few} \times 1000 \km$ the jets cannot penetrate the gas any more because of their jittering. At those large radii the ram pressure of the in-falling gas can be neglected due to its slow inward velocity. 
(3) The jets of each pair of jets deposit their energy inside the star at ${\rm few} \times 1000 \km$ via shock waves, and form two hot bubbles. The bubbles from several pairs of jets merge to form a large bubble that fills most of the volume (see 3D hydrodynamical simulations in \citealt{PapishSoker2014}). This large bubble accelerates the rest of the star and lead to explosion. 
The last assumption allows us to use a spherically symmetric approximation for the expansion of the bubble at ${\rm few} \times 1000 \km $ to $\approx 10^4 \km$. 

The pre-shock (up-stream) ambient density profile at radius $r>R_s$, is (see \citealt{PapishSoker2012} for the usage of this profile)
%with the scaling from \citet{wilson1986} and \citet{Mukami2008} (see Paper 1)
\begin{equation}
\rho_s(r) =  A r^{\beta} =
 1.3 \times 10^{10} \left( \frac {r}{100 \km} \right)^{-2.7} \g \cm^{-3}, \quad 30 \la r \la 10^4 \km
\label{eq:rhos}.
\end{equation}
The spherical flow obeys the following mass, momentum, and energy conservation equations \citep{Volk1985}
\begin{equation}
\frac{dM_s}{dt}=4 \pi R_s^2 \rho(R_s) \dot R_s; \qquad 
\frac{d}{dt} \left( M_s \dot R_s \right ) = 4 \pi R_s^2 P ; \qquad 
\frac{d}{dt} \left ( 4 \pi R_s^3 P \right ) = \dot E_j - 4 \pi R_s^2 P \dot R_s,
\label{eq:mass}
\end{equation}
%% \begin{equation}
%% \frac{d}{dt} \left( M_s \dot R_s \right ) = 4 \pi R_s^2 P \dot R_s,
%% \label{eq:momentum}
%% \end{equation}
%% \begin{equation}
%% \frac{d}{dt} \left ( 4 \pi R_s^3 P \right ) = \dot E_j - 4 \pi R_s^2 P \dot R_s,
%% \label{eq:energy}
%% \end{equation}
where $P$ is the pressure inside the bubble.
The energy inside the bubble includes the thermal energy of the gas and the radiation energy, but neither neutrino losses nor nuclear reactions.
The reason is that we here consider shock radii ($R_s > 1000 \km$) at much larger distances than the small radii ($\approx 100 \km$) where these processes are important.  

This is adequate for the present purpose. 
%We neglect losses by neutrinos (see Paper 1) and energy production and sink from nuclear reactions.

As long as the jets are active with power of $\dot E_j$, the solution to equations (\ref{eq:mass}) is a self-similar solution 
\begin{equation}
R_s(t) = R_0 t^\alpha.
\label{eq:Rst1}
\end{equation}
where 
\begin{equation}
\alpha = \frac{3}{\beta+5}=1.3, \qquad 
R_0 =  \left[ \frac{(\beta+3)(\beta+5)^3}{12 \pi (2 \beta + 7)(\beta+8)}
\frac{\dot E_j}{A} 
\right] ^{1/(\beta+5)}  = 5.2\times 10^8 
\left( \frac {\dot E_j}{10^{51} \erg \s^{-1}} \right) ^{1/(\beta+5)} \cm ,
\label{eq:alpha}
\end{equation}
and where the numerical values are for the same $\beta$ and $A$ from equation (\ref{eq:rhos}). 
A short time of $\Delta t_d \simeq 0.1 \s$ after the jets launching process ceases, energy injection to the bubble ends. The time delay comes from the jets'
crossing time from the NS to the bubble. At that moment the self-similar solution no longer holds, and \cite{PapishSoker2012} present a numerical solution. For our purposes it is sufficient to consider the first second or so. 
 
Substituting equations (\ref{eq:alpha}) in equation (\ref{eq:Rst1}) gives for the shock radius (for the density profile we are using here) 
\begin{equation}
R_s(t) = 
5.2\times 10^8 
\left( \frac {\dot E_j}{10^{51} \erg \s^{-1}} \right) ^{0.43} 
\left( \frac {t}{1 \s} \right)^{1.3} \cm.
\label{eq:Rst2}
\end{equation}
Like \cite{PapishSoker2012}, we scale this equation with the typical activity time period of jets in the jittering jets explosion mechanism, a typical CCSN explosion energy, and the typical radius from which the mass is expelled (or about somewhat smaller at $r \simeq 3000 \km$; \citealt{PapishSoker2012}). Namely, gas at radii smaller than $r \simeq 3000 \km$ fall onto the newly born NS. 

For a mass of $M_\ast = 1.5 M_\odot$ inner to a radius of $r=5000 \km $ the free fall time is $t_{\rm ff}(5000)=0.9 \s$.  
The expansion time for the parameters we use here is about equal to the free fall time at the relevant radius. 
Jets with a much higher power would reach this radius at a shorter time, and would terminate the accretion at an early time. The higher power and the shorter time give about the same total explosion energy. Jets with a much lower power would not stop accretion after one second, and would be active for a longer time, until they have enough energy to expel the core.   
  
We can compare the energy of the jets in this example, $E_{\rm jets} (1 \s) = \dot E_j t \simeq 10^{51} \erg$, to the binding energy of the gas.  
For the above parameters the gravitational energy of the gas is 
\begin{equation}
E_{\rm grav} = -\int^{r \gg R_{\rm in}}_{R_{\rm in}} 
4 \pi r^2 G \rho_s(r) M(r) dr 
\simeq 
- 3 \times 10^{50}
\left( \frac {M_\ast}{1.5 M_\odot} \right)
\left( \frac {R_{\rm in}}{5000 \km} \right)^{-0.7}
\erg
\label{eq:Egrav}.
\end{equation}
The binding energy is somewhat smaller than the absolute value of $E_{\rm grav}$ because of internal energy of the gas. 
 
Let us consider the feedback effect for this particular example. 
If we consider an earlier time, say $t=0.5 \s$ in this example, then the shock would be at a radius of about $2000 \km$. At smaller radii the gravitational energy magnitude will be much larger, about $\vert E_{\rm grav} \vert  \simeq 5.7 \times 10^{50} \erg > E_{\rm jets} (0.5 \s)$. The jets cannot halt accretion. 
 In this example, at about $t \simeq 1 \s$ the jets have enough energy to expel the core, halt accretion, and terminate themselves off.
 The point is that this takes place at $E_{\rm jets} \simeq {\rm {few}} \times E_{\rm bind}$
 
\label{lastpage}
\end{document}